# PRISM: A MATLAB-Based Application for Structured Probe Data Management and Visualization in Tokamak Diagnostics

Priyanka Verma[1,*], Subhojit Bose[1], Harshita Raj[1,2], Joydeep Ghosh[1,2]

[1]Institute For Plasma Research, Gandhinagar-382428, Gujarat - India
[2]Homi Bhabha National Institute, Training School Complex, Anushaktinagar, Mumbai 400094, India

*Email: priyanka.verma@ipr.res.in

## Abstract

The successful operation of tokamak experiments requires accurate documentation, tracking, and visualization of diagnostic instruments, particularly electrical probes. Traditionally, this metadata is maintained manually through handwritten logbooks or semi-digital spreadsheets, leading to inefficiencies and human errors. In response to these challenges, we present PRISM (Probe Registration and Information System for Monitoring)—a MATLAB-based application developed using App Designer. PRISM provides a graphical user interface (GUI) that facilitates structured probe registration, metadata storage, and both 2D and 3D spatial visualization in tokamak geometries. Tested with data from the ADITYA-U tokamak, PRISM helps users enter information accurately, retrieve metadata easily, and visualize probe setups. The tool is built in a flexible way, is not limited to a specific setup, and could potentially support future developments such as digital twins and real-time control systems.

## 1. Introduction

Magnetic confinement fusion research in tokamak devices depends critically on precise and reliable diagnosis of plasma. Among the most fundamental diagnostics, electrical probes offer direct insights into essential edge plasma parameters such as electron density, temperature, potential profiles, and heat flux.

Despite their utility, the management of probe-related metadata—such as installation timelines, spatial configurations, channel mappings, and operational status—has historically relied on physical logbooks or limited-access spreadsheets. These conventional methods are inherently manual, error-prone, and ill-suited to the demands of modern, data-intensive fusion experiments, particularly in multi-user environments or long-duration campaigns.

As fusion experiments become more complex and increasingly collaborative, research facilities have started adopting structured approaches for managing experimental data and associated metadata. At the TEXTOR tokamak, for example, the Experiment Management System (TEXEMS) was introduced to support remote collaboration and organized experiment planning. It combines digital logbooks, planning tools, and metadata capture in a unified framework [1]. Similarly, the MAST tokamak developed the Integrated Data Access Management (IDAM) system to provide consistent access to a broad range of both legacy and new data formats. IDAM also includes built-in mechanisms for signal-level metadata correction and the creation of composite signals [2].

In the case of the COMPASS tokamak, the COMPASS DataBase (CDB) was designed as a scalable and integrated solution for data acquisition, storage, and post-processing. CDB ensures metadata accuracy, uses HDF5-based storage, supports cross-platform access, and offers automatic version control. It also allows seamless integration with various data acquisition systems using language-independent interfaces, streamlining post-shot data workflows [3].

Across the wider scientific community, there is growing recognition of the need for reliable and well-structured data handling methods. The FAIR principles—Findability, Accessibility, Interoperability, and Reusability—serve as a framework to ensure that data can be efficiently located, understood, and reused by both people and computational tools [4]. These principles now play a growing role in the design of fusion data infrastructures. An example is FAIR-MAST, a modern platform for the MAST device that supports public metadata indexing, scalable access via APIs, and integration with Python-based tools, making it suitable for large-scale machine learning and AI applications [5].

Parallel efforts are ongoing in large-scale projects like ITER, where the Data Acquisition Management System (DAMS) has been developed to handle real-time storage and retrieval of extensive datasets, particularly during prolonged DC operations. DAMS uses technologies such as MDSplus and efficient data compression methods to ensure uninterrupted acquisition, metadata handling, and user interaction throughout experiments [6].

Building on this momentum, Greenwald et al. (2012) introduced a structured metadata catalog to help organize and manage fusion simulation data [7]. Their work underlined the importance of using standard metadata formats to make data easier to find, share, and analyze across different research centers. This idea is also relevant for experimental diagnostics, where keeping detailed and accurate records of probe information—such as positions, usage history, and calibration—is important for ensuring reliable data and supporting further analysis or simulation work.

In parallel, the stellarator community has also recognized the importance of digital documentation. At Wendelstein 7-X, an electronic logbook was created to support experiment documentation and collaborative research. It integrates automatically generated logs from the control software with user-provided metadata, enriched through tags and interactive features. Accessible via web interfaces and REST APIs, it has become a central tool for the large international W7-X team [8]. Similar progress has been made in linear plasma devices. For example, in the Large Volume Plasma Device (LVPD), an electronic record-keeping system was developed to facilitate remote participation and structured data documentation, enabling consistent diagnostic records and collaborative work independent of geographical constraints[9].

To build upon these efforts and address the specific needs of tokamak probe diagnostics, we introduce PRISM (Probe Registration and Information System for Monitoring)—a MATLAB-based tool designed for organizing, visualizing, and tracking diagnostic probes used in fusion experiments [10]. PRISM allows users to enter, save, and look up probe details through a user-friendly interface. It also supports both ongoing monitoring and looking back at past data. In addition, the tool offers simple 2D and 3D views of locations of probes inside the tokamak, helping users understand the setup more clearly.

As experimental fusion devices grow increasingly complex, the transition from scattered, error-prone practices to structured, more dependable digital tools become essential. PRISM addresses this shift by offering a centralized, user-friendly platform that ensures metadata consistency, traceability, and ease of access—core requirements for reliable diagnostics and collaborative research. Furthermore, PRISM aligns with emerging trends in fusion science, including the development of digital twin frameworks and automated plasma control systems, where accurate geometric and operational metadata are prerequisites for real-time simulation and decision-making.

PRISM offers a structured and adaptable framework for managing probe metadata, establishing a foundation for future integration with these advanced systems. In addition to supporting ongoing experimental operations, the application helps new researchers get started more easily by simplifying the often-confusing naming systems and older forms of documentation used in diagnostics. With a user-friendly interface, PRISM reduces the effort associated with metadata handling and contributes to a more accessible, consistent, and less error-prone research environment.

## 2. System Design and Requirements

The PRISM application has been developed using MATLAB R2024a, employing the App Designer framework to ensure a structured graphical user interface and modular backend functionality. It is designed to operate in both local execution and remote deployment environments, providing flexibility in how users can access and interact with the system.

For standalone deployment, PRISM is compiled using MATLAB's Application Compiler, enabling it to run on systems without a full MATLAB installation through the MATLAB Runtime environment. While MATLAB Runtime support for standalone applications has been available in earlier releases, PRISM specifically uses the R2024a version to leverage the latest performance improvements, toolbox updates, and compatibility with the current MATLAB ecosystem. This approach ensures end users can operate the application without needing individual MATLAB licenses or installations, thereby improving accessibility.

When broader access and centralized control are required, PRISM can also be hosted via MATLAB Web App Server. Building on the same core codebase as the standalone version, this server-based deployment enables multi-user interaction through standard web browsers without client-side installation. The server environment must have MATLAB R2024a installed along with the MATLAB Compiler and Web App Server components to manage application execution, handle requests, and facilitate data interaction.

The underlying data management system relies on Microsoft Excel files (.xlsx format), which serve as the central database for input, retrieval, and archival of experimental records. These files are stored on a shared network drive to facilitate centralized access, consistency of records, and concurrent usage by multiple authorized users. The application is designed to interact dynamically with this backend, allowing for seamless integration of new data without requiring manual intervention. The PRISM software is optimized for deployment within an institutional IT infrastructure, supporting secure and stable network access. Its cross-platform accessibility is ensured through browser-based interaction, with no specialized hardware or operating system dependencies on the client side. Overall, the design emphasizes modularity and scalability to accommodate evolving research needs and institutional usage scenarios.

The workflow of PRISM is given in Fig.1.

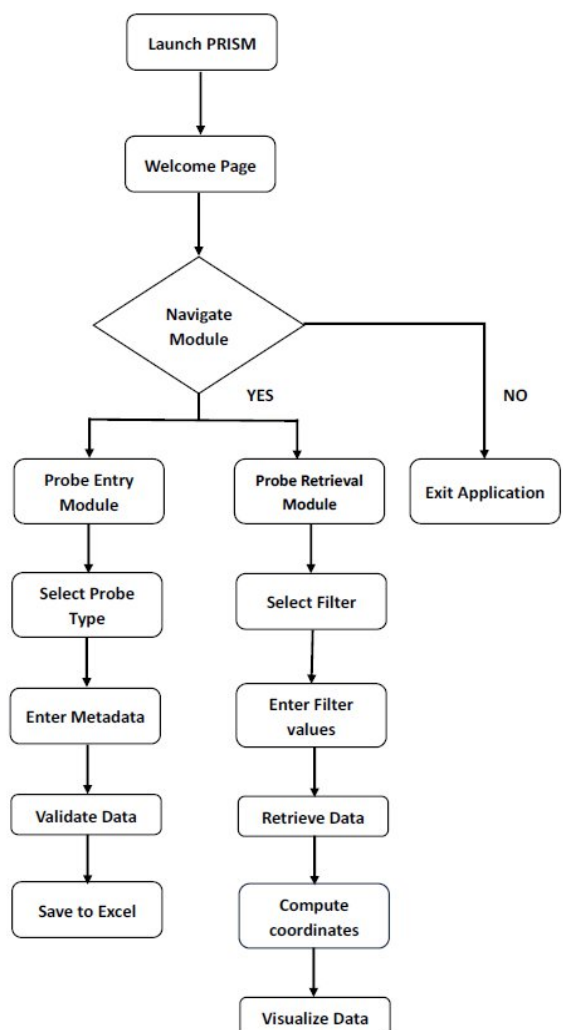

*Fig. 1) PRISM Workflow and Functionality*

## 3. Software Architecture

### 3.1 App Designer Framework

PRISM is built using MATLAB App Designer's Design View for drag-and-drop UI layout and Code View for defining app behavior using MATLAB's object-oriented capabilities. Each app developed in this framework inherits from matlab.apps.AppBase.

PRISM's interface includes tabs, buttons, dropdowns, tables, and plotting components. The modular structure enables clear separation between probe registration, data retrieval, and visualization functionalities.

### 3.2 Data Handling and Visualization

The application uses Excel files as the backend database, with each probe type assigned a dedicated worksheet. Probe entries are logged with metadata fields including probe name, date, shot number, shot duration, channel number, port type, position, angular coordinates and other experimental details. The database is automatically sorted by date, and built-in error handling prevents entry of invalid, incomplete or duplicate data.

Upon retrieval, based on the selected filter, probe data are fetched and visualized as follows:

- 2D poloidal and top-view plots illustrate the radial and angular distribution of probes.
- 3D toroidal plots depict probe positions by transforming toroidal coordinates $(r,\theta,\phi)$ into Cartesian space.

The visualization aids in spatial verification of diagnostic placement and supports correlation with plasma position data during experiments.

## 4. Workflow and Methodology

PRISM consists of three primary functional windows, each serving a specific purpose in the workflow: the Welcome Page, Probe Entry Module, and Probe Retrieval Module.

### 4.1 Welcome Page

Upon launching PRISM, users are greeted with a centralized Welcome Page that serves as the main entry point to the application. This interface is designed to be both user-friendly and informative, making it easy for both new and experienced users to navigate the software efficiently. The Welcome Page consists of two main sections that guide users through essential features and usage guidelines.

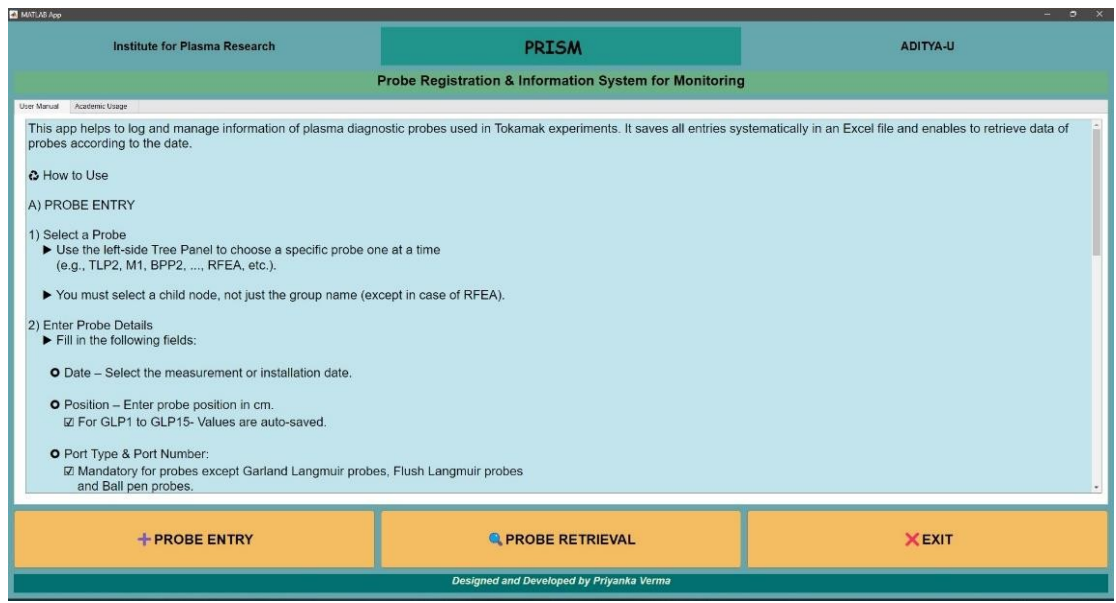

*Fig. 2) Welcome page of PRISM application showing the instructions in user manual tab with buttons- Probe Entry, Probe Retrieval, Exit.*

The User Manual tab offers a detailed overview of PRISM's core functionalities. It provides step-by-step instructions for entering new probe data, retrieving historical records, and interpreting spatial visualizations. This section is especially helpful for first-time users, supporting a smooth onboarding experience through clearly outlined procedures.

The Academic Usage tab outlines licensing terms, usage conditions, and proper citation practices for scholarly publication. It includes a link to the software's Zenodo repository, ensuring reliable access for long-term use, citation, and reproducibility in academic research.

In addition to these resources, the Welcome Page includes clearly labelled navigation buttons that allow users to move directly to the Probe Entry module for submitting new data, the Probe Retrieval module for accessing and visualizing archived records, or to securely exit the application. The welcome page of PRISM is shown in Fig. 2.

**4.2 Probe Entry Module**

This module enables users to register new diagnostic probes. It presents a structured tree view of probe categories, from which the user selects the appropriate type (e.g., Langmuir probe, Mach probe etc.). Once selected, the user must populate metadata fields such as installation date, shot number and duration, port type, channel number, spatial coordinates and other experimental details.

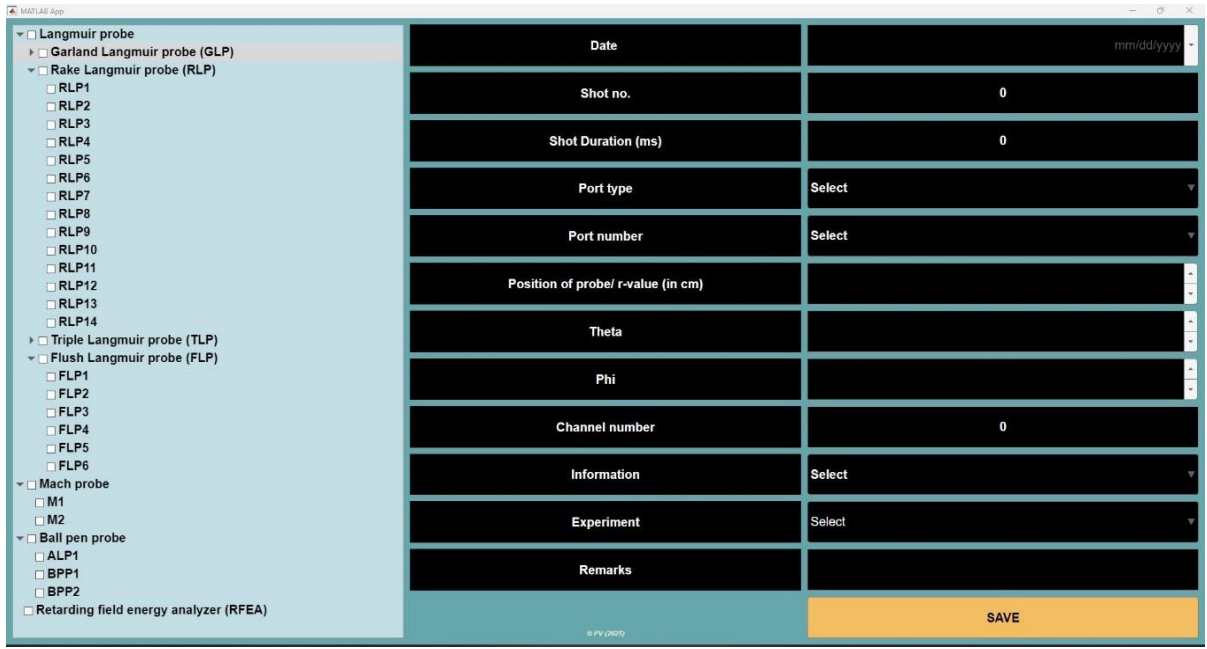

*Fig. 3) Probe entry window with tree structure and components like tabs, spinners, text boxes for metadata entry*

The interface includes validation routines to ensure that all mandatory fields are completed and follow proper data formats. Pressing the SAVE button commits the information to the Excel backend, updating the appropriate worksheet for the selected probe category. The module automatically checks for duplicate entries and incorrect formatting to prevent database inconsistencies. Probe Entry Window is shown in Fig.3.

**Table 1. Components of Probe Entry Module**

| Component | Type (UI Element) | Description / Functionality |
|---|---|---|
| CheckBoxTree | TreeNode with checkboxes | Displays all probes in a hierarchical arrangement; probes of the same type are grouped under one node (Fig. 3, left panel). |
| Date | DatePicker | Allows user to select the date of probe entry. |
| Shot No. | Numeric Edit Field | Enter the shot number for the selected date. |
| Shot Duration (ms) | Numeric Edit Field | Duration of plasma shot in milliseconds for the selected shot number. |
| Port Type | DropDown Menu | Select port type: *Radial, Top, Bottom*. |
| Port Number | DropDown Menu | Select the port number where the probe is located. |
| Position of Probe (r) | Numeric Spinner | Enter radial position of probe's tip in cm; input restricted to allowed range. |
| Theta ($\theta$) | Numeric Edit Field | Enter poloidal angle of probe's location (degrees). |
| Phi ($\varphi$) | Numeric Edit Field | Enter toroidal angle of probe's location (degrees). |

| Component | Type (UI Element) | Description / Functionality |
|---|---|---|
| Channel Number | Numeric Edit Field | Enter channel number where probe signal is stored. |
| Information | DropDown Menu | Select type of stored information: *Monitor, Temperature, Density, Floating Potential, Other*. |
| Experiment | DropDown Menu | Select experiment type ongoing on the date of entry. |
| Remarks | Text Field | Enter any extra information related to probe entry. |
| Save | Button | Saves the values in Excel file after all the components are filled. |

**4.3 Probe Retrieval and Visualization Module**

In this module, users can retrieve and visualize previously entered probe configurations based on several filtering options. Based on the selected filter, the corresponding input field is displayed—for example, a date selector, experiment dropdown, or numeric input box for position or duration. When filtering by radial position, users are additionally required to enter the angle corresponding to the probe's toroidal location.

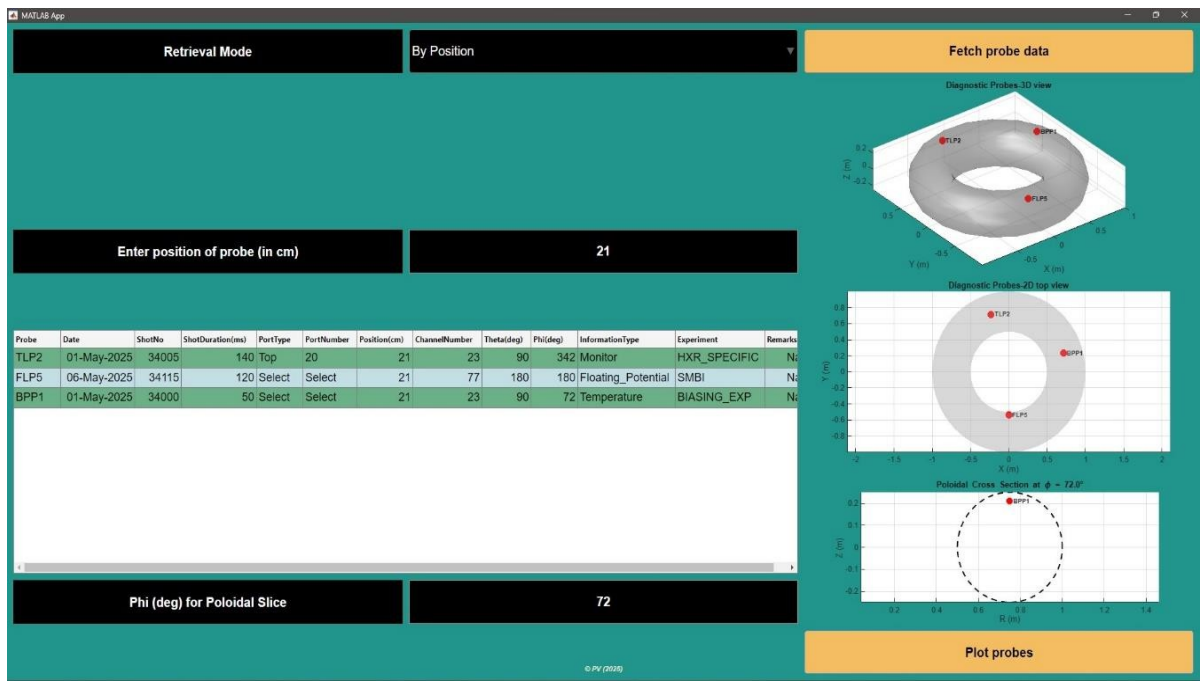

*Fig. 4) Probe Retrieval Window showing sample data filtered by position of probes with 2-D and 3-D views of the locations of probes.*

Upon entering the desired values as per the selected filter option, PRISM fetches the corresponding entries by aggregating probe metadata from all relevant worksheets within the Excel database. The retrieved data are displayed in a structured tabular format within the app, with column headers consistent with those defined in the original Excel sheets.

Subsequently, the spatial coordinates of the probes are computed by transforming the toroidal coordinates (r,θ,ϕ) into Cartesian coordinates (x,y,z) using standard toroidal-to-Cartesian conversion relations. Adjustments are applied to account for the directionality of angular parameters—i.e., clockwise or counterclockwise orientation of θ and ϕ—and to align the calculated positions with the defined reference axes of the tokamak geometry.

The processed coordinates are visualized using MATLAB's built-in plotting capabilities through a dual-view representation that enhances spatial understanding. A three-dimensional toroidal geometry provides an accurate spatial depiction of probe placement within the vessel. In addition, two-dimensional projections are used to offer complementary perspectives: a top view emphasizes the radial and angular distribution of probes along the toroidal axis, while a poloidal view primarily highlights the radial positioning of probes within a cross-sectional plane of the torus.

This combined visual display allows researchers to confirm probe placement and better understand spatial relationships within experimental data. The intuitive layout and responsive interface make these features easy to use, even for those with limited technical experience. Probe retrieval window with sample data is shown in Fig.4.

**Table 2. Components of Probe Retrieval Module**

| Component | Type (UI Element) | Description / Functionality |
|---|---|---|
| Retrieval Mode | DropDown Menu | Select retrieval filter: *By Date, By Shot No., By Position, By Shot Duration, By Experiment*. |
| Select Date | DatePicker | If retrieval mode = *By Date*, select the date for retrieving probe data. |
| Enter Shot No. | Numeric Edit Field | If retrieval mode = *By Shot No.,* enter the shot number. |
| Enter Position of Probe | Numeric Edit Field | If retrieval mode = *By Position*, enter radial probe tip position (cm). |
| Phi for Poloidal Slice | Numeric Edit Field | Enter toroidal angle of probe's location for 2-D plotting. |
| Enter Shot Duration | Numeric Edit Field | If retrieval mode = *By Shot Duration*, enter plasma shot duration. |
| Experiment | DropDown Menu | If retrieval mode = *By Experiment*, select experiment type. |
| Fetch Probe Data | Button | Click after selecting criteria and filling inputs to retrieve the relevant probe data. |
| Plot Probe Data | Button | Once probe data is retrieved, click to generate 2-D and 3-D plots of probe locations. |

## 5. Results and Discussion

PRISM has been thoroughly tested using diagnostic datasets from the ADITYA-U tokamak, demonstrating noticeable improvements in how probe metadata is managed over several experimental campaigns. By automating tasks that were traditionally manual—such as data entry and retrieval—it has significantly streamlined workflows and reduced time spent on routine processes. This shift has also improved the consistency and accuracy of metadata while minimizing duplication and errors commonly found in manual record-keeping systems.

The tool enforces a structured data entry format that helps maintain uniformity across users and experimental runs, thereby reducing the chance of inconsistencies or user-related errors. Its built-in retrieval function supports quick access to historical probe configurations, which is especially helpful for comparing experimental results over time and conducting detailed post-shot analysis.

A key feature of PRISM is its ability to generate 2D and 3D visualizations of probe placement within the tokamak geometry. The visualization aids in spatial verification of diagnostic placement and supports correlation with plasma position data during experiments. These capabilities simplify the interpretation of spatial information, assist in identifying configuration anomalies early, and enhance the quality of experimental planning and post-analysis. By improving both the accuracy and accessibility of probe metadata, PRISM contributes meaningfully to the reliability and efficiency of diagnostic workflows in fusion research.

## 6. Conclusion and Future Work

This paper presents PRISM, a MATLAB-based application designed for the structured registration, visualization, and management of diagnostics in tokamak environments. Developed within the App Designer framework and rigorously validated with experimental data from the ADITYA-U tokamak, PRISM offers a modular and extensible platform tailored to the evolving requirements of fusion diagnostics.

Future developments will focus on enhancing the application's capabilities through integration with real-time plasma control and monitoring systems, enabling cloud-based multi-user access with concurrent editing functionality, and achieving compatibility with standard data acquisition hardware such as PXI and CAMAC platforms. Additionally, expansion to support a broader range of diagnostic tools—including bolometers, spectroscopy arrays, and microwave interferometers—is planned to further unify the management of tokamak diagnostic infrastructure.

Another important direction involves the incorporation of machine learning algorithms to automate the validation of probe signals—specifically to assess whether the data stored in corresponding probe channel numbers is meaningful or corrupted—thereby reducing manual verification overhead and improving the reliability of analysis-ready datasets.

By systematically addressing persistent challenges in probe metadata management, PRISM represents a significant step toward the digital transformation of experimental fusion research. It facilitates improved reproducibility, data integrity, and operational efficiency, thereby supporting more robust and transparent tokamak diagnostic workflows.

## 7. Software Availability

The PRISM application (Probe Registration and Information System for Monitoring) is archived and accessible via the Zenodo repository: https://doi.org/10.5281/zenodo.15334041

PRISM is distributed under the Creative Commons Attribution-NonCommercial-ShareAlike 4.0 International (CC BY-NC-SA 4.0)[1] license, permitting academic and non-commercial use.

**CRediT authorship contribution statement**

**Priyanka Verma:** Conceptualization, Methodology, Software, Validation, Data curation, Writing – original draft, Visualization. **Subhojit Bose:** Conceptualization, Writing – review & editing. **Harshita Raj:** Conceptualization, Supervision. **Joydeep Ghosh:** Supervision.

## Acknowledgments

The author gratefully acknowledges Soumitra Banerjee and Injamul Hoque for their valuable feedback throughout the iterative development of the PRISM application, including suggestions for feature enhancements and support with testing. The author also appreciates the constructive discussions and thoughtful comments provided by Dr. Shwetang N. Pandya and Dr. Ritu Dey. The valuable support of the ADITYA-U tokamak team during the software validation phase is also gratefully acknowledged.

---

[1]**License:** Creative Commons Attribution-NonCommercial-ShareAlike 4.0 International (CC BY-NC-SA 4.0) https://creativecommons.org/licenses/by-nc-sa/4.0/